\documentstyle[preprint,aps]{revtex}
\draft
\begin{document}
\begin{title}
{Weak universality in the two dimensional randomly
disordered three-state Potts ferromagnet}
\end{title}
\author{Jae-Kwon Kim}
\address
{Department of Physics, University of California,
Los Angeles, CA 90095-1547}
\maketitle

\begin{abstract}
For the two dimensional randomly disordered three-state  Potts
ferromagnet, we find numerically that the critical exponent $\eta$
unchanges with the degree of disorder while those
of the correlation length ($\nu$) and
magnetic susceptibility ($\gamma$)
increase with it continuously.
We discuss some consequences of the finding.
\end{abstract}

The effect of random quenched disorder to the critical
behavior of a ferromagnetic system
has been a subject of intensive studies.
According to a heuristic argument by Harris\cite{HAR},
that has been widely accepted over two decades,
the disorder induces a new universality class only when
the critical exponent of the specific heat ($\alpha_{p}$)
in the corresponding undisordered system  is positive.

A more rigorous approach\cite{CHAY} claims
that in a D-dimensional disordered
system obeys
\begin{equation}
\nu \ge 2/D.    \label{eq:ine}
\end{equation}
The proof of Eq.(\ref{eq:ine}), however, presumes the existence
of so-called {\it finite-size scaling correlation length}
whose relation to the intrinsic correlation
length $\xi$ is by no means obvious,
defined in terms of probability.
Note that Eq.(\ref{eq:ine}) combined with
hyperscaling relation, $\alpha_{p}=2-D \nu_{p}$ (with
$\nu_{p}$ denoting the value of $\nu$ in the corresponding
undisordered system), implies that
for a disordered system with $\alpha_{p}>0$
$\nu$ {\it jumps} discontinuously from
the undisordered value $\nu_{p}$.
To be more specific, random disorder is supposed to
induce a different scaling region near criticality
while the rest of the scaling region is equivalent to
that of the undisordered system
(crossover from the undisordered  to disordered).

The 2D random bond disordered Ising ferromagnet is
unique in that its continuum limit is found to be
a certain tractable field theoretic model, i.e.,
$O(N)$ Gross-Neveu model in the limit $N \to 0$.
Accordingly it is predicted that the value of
$\eta$ in this system does not vary with the strength
of weak disorder\cite{SHA}, which  has now been confirmed
by various numerical methods even for strongly disordered case
\cite{WANG,SLA,PAT,KUHN,KIMI}.
Albeit $\alpha_{p}=0$, recent extensive numerical
studies\cite{PAT,KUHN,KIMI}
find that both $\nu$  and $\gamma$ change with disorder,
in disaccord with Harris criterion.

Most numerical studies
on the 3D random site diluted Ising ferromagnet,
where $\alpha_{p}  \simeq 0.1$,
report evidences for new universality class\cite{NOV,HOL,BRA}.
Still is it not clear whether Eq.(\ref{eq:ine}) is indeed
satisfied or not\cite{HOL};
nor is it clarified whether
critical exponents change continuously with
the strength of disorder or jump to certain fixed
values regardless of it.
An accurate estimate of $\eta$ is far more
difficult, for it is very sensitive to the choice of
the critical point.
Generally speaking, current understanding on generic
randomly disordered ferromagnet is rather limited
theoretically, numerically, and experimentally.

In this Letter we report new finding of the extensive
Monte Carlo studies on the 2D random bond disordered
three-state Potts ferromagnet.
The technical difficulty is
significantly lightened in this system, since the
exactly known value of $\alpha_{p}$ is large enough
($\alpha_{p}=1/3$) and the exact critical point can be
calculated for certain realizations of disorder.
Thus we obtain clean numerical evidence
that the value of $\eta$ unchanges with disorder
while both $\nu$ and $\gamma$ increase continuously.
With the very same results found in the 2D disordered
Ising ferromagnet\cite{KIMI},
we conjecture that those features are generic
in any randomly disordered ferromagnets.

The Hamiltonian of a $q$-state Potts ferromagnet
with quenched random interaction (bond) can be written as
\begin{equation}
H= -\sum_{<x,y>} J_{xy} \delta_{\sigma_{x} \sigma_{y}},
\end{equation}
where the spin at site $x$,
$\sigma_{x}$, can take on the values 1,...,$q$ ($q=3$ in this work),
$\delta$ is the Kronecker delta
function, and the sum is over all the nearest-neighbor
pairs on the lattice system.
The coupling $J_{xy}$ for each link $<x,y>$ is selected
randomly from two positive (ferromagnetic) values $J$
and $J^{\prime}$
with probability $p$ and $1-p$ respectively.

For $p=1/2$ the system is self-dual with its self-dual
point given by\cite{DOM}
\begin{equation}
[\exp(J)-1][\exp(J^{\prime})-1]=q
\end{equation}
We fix $J=1$ and $p=1/2$ without any loss of generality, and
consider four different values of $J^{\prime}$, i.e.,
$J^{\prime}=$1.0 (undisordered case), 0.9, 0.5, and 0.25.
Accordingly, the inverse self-dual temperature (critical temperature)
is given by $\beta_{c}=\ln(\sqrt{3}+1)$, 1.0588...,
$2\ln(2)$, and 1.8235..., respectively
for $J^{\prime}$=1.0, 0.9, 0.5, and 0.25.
The values of the critical exponents for the undisordered case
are exactly known\cite{WU}, e.g., $\nu_{p}=5/6$ and $\gamma_{p}=13/9$
(hence $\eta_{p}=4/15$).

Our raw-data for each $J^{\prime}$ are obtained by choosing
a realization of random $J^{\prime}$, then running Monte Carlo
simulations in Wolff's single cluster algorithm with
periodic boundary conditions imposed on square lattice.
Details of our single cluster algorithm for Potts model
will be presented in a longer paper.
For each realization, measurements were taken
over 10 000 configurations
each of which is separated by 6-24 one cluster updatings
according to  autocorrelation time. The mean values over
all the different realizations are our final data.
The numbers of different realizations
are typically 30-60, 150-200, and 250-350 respectively
for $J^{\prime}=0.9$, 0.50 and 0.25.
Larger number of realizations are
required for smaller value of $J^{\prime}$ due to
wilder fluctuations among different realizations.

Our magnetic susceptibility and correlation length on
a square lattice of linear size L are defined as
\begin{eqnarray}
\chi_{L} &=& \tilde{G}(0), \\
\xi_{L}&=& {1 \over \sin(2\pi/L) } \sqrt{\tilde{G}(0)/\tilde{G}(2\pi/L)-1},
\end{eqnarray}
where the two-point correlation function $G(x)$ and its
Fourier transformation
$\tilde{G}(k)$ are respectively defined as
$G(x) \equiv <\delta_{\sigma_{0} \sigma_{x}}-1/q>$ and
$\tilde{G}(k)\equiv \sum_{x}G(x)\exp(kx_{1})$ with $x_{1}$
denoting the first component of $x$.

In order to determine the value of $\eta$
we measured $\chi_{L}$ at each critical point (self-dual point)
by varying L from 20 to 120.
According to the standard theory of finite size scaling,
$\chi_{L}$ at criticality satisfies
\begin{equation}
\chi_{L} \sim L^{2-\eta},
\end{equation}
where $\eta=2-\gamma/\nu$.
Our results are summarized in Figure(1) showing undoubtedly
that the value of $\gamma/\nu$ remains unchanged irrespective
of the value of $J^{\prime}$. The $\chi^{2}$ fits yield
$\eta$=0.266(3), 0.264(3), 0.267(8), 0.268(10), respectively for
$J^{\prime}$=1.0, 0.9, 0.5, and 0.25, to be compared with the exact
value 0.266... ($\chi^{2}/N_{DF} <0.5$ for all the $J^{\prime}$).

The value of $\gamma$ is determined by the
$\chi^{2}$ fit of bulk $\chi$ data measured at various
values of the inverse temperature $\beta$.
The data with $\xi(\beta) \ge 5$ are considered in the fit
to eliminate possible correction to the scaling
\begin{equation}
\chi \sim t^{-\gamma} \label{eq:fit},
\end{equation}
where $t\equiv (\beta_{c}-\beta)/\beta_{c}$.
We used $\xi_{L}$ to monitor the finite-size effect
in our measurements of bulk $\chi$;
that is, the condition $L/\xi_{L} \ge 8$
was always imposed for the extractions of
our bulk data (see Figure(2)).
The largest values of $\xi$ are roughly $20 \sim 30$
for all the values of $J^{\prime}$, using L up to L=240.

Our procedure for the undisordered case
yields $\gamma$=1.442(3) in excellent agreement with
the exact value $\gamma$=1.444... .
The data for different $J^{\prime}$ are summerized in Figure(3).
The value of $\gamma$ clearly increases with the
strength of disorder: $\gamma$=1.44(1), 1.56(4), and 1.84(1)
respectively for $J^{\prime}$=0.9, 0.5, and 0.25
($\chi^{2}/N_{DF} <1$).
Also we do not find any indication of crossover,
as  in the 2D randomly disordered Ising
ferromagnets\cite{PAT,KIMI}. In other word,
the presence of random quenched disorder
affects the entire scaling regime (Figure(3)).

For the weakly disordered case ($J^{\prime}=0.9$)
the  value of $\gamma$ thus extracted is virtually the
same value as in the undisordered case.
We do not take this as an evidence that their values
are strictly the same in those two systems; it rather
appears that $\gamma$ increases indistinguishably mildly
for a very weakly disordered system.

Eq.(\ref{eq:ine}) combined with
our finding that $\gamma/\nu$ remains a constant
implies that $\gamma \ge 26/15$
in the two dimensional randomly disordered three-state Potts
ferromagnets, in obvious disaccord with our results for
$J^{\prime}=0.9$ and 0.5.

Our results contrast with some previous numerical results
in the 3D disordered Ising ferromagnet\cite{CHO}\cite{LAN},
but are consistent with the results in Ref.\cite{BRA} where
the critical exponent of the magnetization was found to
vary continuously.
The variant conclusions of the previous numerical studies
on the 3D disordered Ising ferromagnet are perhaps
due to some technical reasons: Without measuring
correlation length, a priori, the finite size effect
in the measurement of bulk data can hardly be monitored, nor
can there be a criterion to tell whether or not a data
point $(\beta,\chi(\beta))$ is in the scaling regime.
Moreover, without knowing
precise critical point one usually needs very broad range
of the data for a reliable extraction of the parameters
from a fit to Eq.(\ref{eq:fit}).
Our current investigation is free of all the drawbacks.

Our results bear striking resemblance to the results
of the 2D random-bond disordered Ising ferromagnet\cite{KIMI}:
There the data up to $\xi \simeq 540$ are analyzed for the
similar values of $J^{\prime}$  to those here, and it
is found that $\nu$ and $\gamma$ for $J^{\prime}=0.9$
are virtually the same values as
in the corresponding undisordered system while
they are clearly increased for $J^{\prime}=0.25$ and 0.1.
It thus appears that the critical behavior of
a disordered system
is not characterized by the value of $\alpha_{p}$.

Some time ago Suzuki proposed\cite{SUZ}
the concept of {\it weak universality}
to comprise the rigorously known
critical exponents of the eight-vertex model into
the concept of universality.
According to it the value of $\nu$ may change
with the details of microscopic interaction
whereas the value of $\eta$ does not, so that
a (weak) universality class is characterized by the
value of $\eta$, but not by the value of $\nu$ or $\gamma$.
The presence of the random quenched disorder does not
change the symmetry of the system
although microscopic interaction changes.
Thus, in view of the weak universality
the value of $\eta$ remains the same value as
in the corresponding undisordered system whereas
both $\nu$ and $\gamma$ vary.

Some non-trivial examples of the realization of
the same weak universality class would be
the D-dimensional XY and Ising models, for
the Z(2) group in the Ising ferromagnet is
the center group of the O(2) symmetry in the XY model.
It has recently been proved numerically that
the value of $\eta$ in the 2D XY model is indeed 1/4\cite{KIMXY}
(as in the 2D Ising model),
while $\eta=0.031(4)$ and 0.026(3)\cite{BAI}
respectively for the 3D XY and Ising models
being equal to each other within the statistical errors.
We conjecture that the similar pattern of weak universality
should manifest itself even in a disordered system
with $\alpha_{p} <0$: the 3D disordered XY and Ising models
should be in the same weak universality class
as the corresponding undisordered models, albeit
$\alpha_{p}$ is negative (positive) in the 3D undisordered
XY (Ising) models.

To conclude: We have shown that the value of $\eta$ in the random
bond-disordered three-state Potts ferromagnet does not change
with the strength of disorder, while both $\nu$ and $\gamma$ vary
continuously.
Our results disprove so-called semi-rigorous inequality,
but satisfies the concept of weak universality.
The features found in the model
seem to be generic in any disordered ferromagnets irrespective
of the value of $\alpha_{p}$, contrary to the conventional wisdom.

{\bf Figure Captions} \\
Figure(1): $\ln (\chi_{L})$ as a function of $\ln (L)$.
         The slope of a straight line would correspond to $2-\eta$.
         The dotted line represents the case of the exact value of $\eta$
         of the undisordered system. \\
Figure(2): $\chi_{L}/\chi_{\infty}$ as a function of $L/\xi_{L}$
         at certain fixed temperatures.
         For each $J^{\prime}$, $\chi_{L}$ becomes equivalent to its
         bulk value within the statistical errors under the condition
         $L/\xi_{L} \ge 5$. Note that this condition is
         independent of temperature according to the standard theory
         of finite size scaling, that is, e.g.,
         $\chi_{L}(\beta)/\chi_{\infty}(\beta)
         =q_{\chi}(\xi_{L}(\beta)/L)$.\\
Figure(3): $\ln (\chi(t))$ as a function of $|\ln (t)|$. The slope of
         a straight line is equivalent to the value of $\gamma$.
	 Each dotted line represents the best approximation of
         the data obtained from the $\chi^{2}$ fit. It is clear that
         the value of the slope increases with decreasing $J^{\prime}$
	 for $J^{\prime} \le 0.9$, and that for each $J^{\prime}$
         it does not vary with $t$.
\end{document}